\documentclass[12pt]{article}
\usepackage{epsf,amssymb}

\input{pdfuc2.tbl}
\input{pdfuc2.fig}
\input{pdfuc2.pre}

\begin{document}


\begin{titlepage}

\begin{tabular}{l}
\noindent{January 4, 2001}  
\end{tabular}
\hfill
\begin{tabular}{l}
\PPrtNo
\end{tabular}

\vspace{1cm}

\begin{center}
\renewcommand{\thefootnote}{\fnsymbol{footnote}}
{
\LARGE \TITLE
}

\vspace{1.25cm}
{\large  \AUTHORS}

\vspace{1.25cm}

\INST
\end{center}

\vfill

\ABSTRACT                 

\vfill

\newpage
\end{titlepage}

\renewcommand{\thefootnote}{\alph{footnote}}   
\setcounter{footnote}{0}

\tableofcontents
\newpage

\section{Introduction}

\label{sec:Intro2}

All calculations of high energy processes with initial hadrons, whether
within the Standard Model (SM) or exploring New Physics, require parton
distribution functions (PDFs) as an essential input. The reliability of
these calculations, which underpins both future theoretical and experimental
progress, depends on understanding the uncertainties of the
PDFs. The assessment of PDF uncertainties has, therefore, become an
important challenge to high energy physics in recent years.

The PDFs are derived from global analyses of experimental data from a wide
range of hard processes in the framework of perturbative quantum
chromodynamics (PQCD). Quantifying the uncertainties in a global
QCD analysis is far from being a straightforward exercise in statistics.
There are non-gaussian sources of uncertainty from perturbation
theory ({\it e.g.,} higher order and
power law corrections), from choices of parametrization of the
nonperturbative
input ({\it i.e.,} initial parton distributions at a low energy scale), from
uncertain nuclear corrections to experiments performed on nuclear targets,
and from normal experimental statistical and systematic errors. These
sources of error need to be studied individually,
and eventually combined in a systematic way.

We shall be concerned in this paper with uncertainties of PQCD
predictions due to uncertainties of PDFs arising from experimental
measurement errors.
This problem is considerably more complicated than it
appears on the surface.
The reason is that in a \emph{global analysis},
the large number of data points ($\sim 1300$ in our case)
do not come from a uniform set of measurements, but consist of a collection
of measurements from many experiments ($\sim 15$) on a variety of physical
processes ($\sim 5-6$) with diverse characteristics, precision, and error
determination.
The difficulty is compounded by a large number of fitting
parameters ($\sim 16$) which are not uniquely specified by the theory.
Several approaches to this problem have been proposed, with rather different
emphases on the rigor of the statistical method, scope of experimental
input, and attention to various practical complications\ \cite
{ZomerDis,Alekhin,GieleEtal,Botje,Bialek,Ball,RunII}.
Our group has initiated one of
these efforts, with the emphasis on utilizing the full constraints of the
global data \cite{RunII}. This work has motivated a closer examination of
the standard techniques of error analysis, and necessary improvements and
extensions to these techniques,
as applied to a complex real world problem such as
global QCD analysis of PDFs \cite{pdfuc0}.

In this paper we present a detailed analysis of uncertainties of physical
observables due to parton distribution functions, using the Lagrange Multiplier
method proposed in \cite{RunII, pdfuc0}.
This method explores the entire
multi-dimensional parton parameter space, using an effective $\chi^2$ function
that conveniently combines the global experimental, theoretical, and
phenomenological inputs to give a quantitative measure of the goodness-of-fit
for a given set of PDF parameters. ({\it Cf.}\ Sec.~\ref{sec:Global}.)
The method probes directly
the variation of the effective $\chi^2$ along a specific
direction in the PDF parameter space---that of maximum variation of a
specified physical variable.
The result is a robust set of \emph{optimized sample PDFs} (or
``alternative hypotheses'') from which the uncertainty of the physical variable
can be assessed quantitatively without the approximations inherent in the
traditional error matrix approach. For concreteness, we consider the cross
section $\sigma_{W}$
of $W$ boson production at the Tevatron as the archetypal example. %
({\it Cf.}\ Sec.~\ref{sec:Lagrange}.)

The definition of the effective $\chi^2$ function,
and the inputs that go into it, do not permit a direct
statistical interpretation of its numerical value.
To obtain meaningful confidence levels for the optimized sample PDF sets,
it is necessary to conduct a series of likelihood analyses of these sample
PDFs, using all available information on errors for the individual
experiments.
The results from these analyses serve as the basis
to assign an overall uncertainty range on the physical
variable,
and a corresponding tolerance measure for the
effective $\chi^2$ function used in the analysis,
that are consistent with the experiments used in the
current global QCD analysis. %
({\it Cf.}\ Sec.~\ref{sec:quantifying}.)

This method can be applied to any physical variable, or combination of
physical variables, in precision QCD phenomenology. In
Sec.~\ref{sec:MoreExamples} we present results on $W$ production at
the LHC, and $Z$ production at the Tevatron and the LHC.
We compare the uncertainties obtained in
all cases, and comment on previous estimates in the context of these
results. In Sec.~\ref{sec:UpDownPdf} we present parton distribution sets
that are optimized to give high/low values of the $W$ and $Z$ cross
sections, while remaining consistent with current experiments according
to our analysis.

The Lagrange Multiplier method provides a useful tool to test the reliability
of the more traditional method of error propagation via the error matrix
\cite{ZomerDis,Botje,CSCTEQ}, which relies on the quadratic expansion of the
$\chi^2$ function around its minimum.
In a companion paper \cite{Hesse} we perform an in-depth analysis of the
uncertainties of the PDFs in the error matrix approach, using the much
improved numerical method for calculating the Hessian that was
developed in \cite{pdfuc0}.
There we demonstrate how the more specialized Lagrange Multiplier method
can set useful benchmarks for the general purpose error matrix approach.

\section{The Global QCD Analysis}

\label{sec:Global}

We adopt the same experimental and theoretical input as the CTEQ5
analysis \cite{cteq5}: 15 data sets from 11 experiments on neutral-current and
charged-current deep inelastic scattering (DIS), lepton-pair production
(Drell-Yan), lepton asymmetry in $W$-production, and high $p_{T}$ inclusive
jet production processes are used.
({\it Cf.} Table \ref{tbl:DatSet} in Sec.~\ref{sec:quantifying}.)
The total number of data points is $N=1295$.
We denote the experimental data values by
$\{D\}=\{D_{I};\;I=1,\dots,N\}$.
The theory input is next-leading-order (NLO) PQCD, and the
theory value for the data point $I$ will be denoted by $T_{I}$.
The theory depends on a set of parameters
$\{a\} \equiv \{a_{i};\;i=1,\dots,d\}$.
These parameters characterize the nonperturbative
QCD input to the analysis; they determine the initial PDFs
$\{f(x,Q_{0};\{a\})\}$ defined at a low energy scale $Q_{0}$,
below the energy scale of the data, which we choose to be
$Q_{0}=1$\,GeV.
When we need to emphasize that the theoretical values depend
on the PDF parameters we write $T_{I}(a)$ to indicate
the dependence on $\{a\}$.

The parametrization of $\{f(x,Q_{0})\}$ is somewhat arbitrary,
motivated by physics, numerical considerations, and economy.
Another parametrization might be employed,
and differences among the possible parametrizations are in principle a
source of theoretical uncertainty in their own right.
For most of this study we focus on a single parametrization,
but we comment on the effect of changing the parametrization
at the end of Sec.~\ref{sec:quantifying}.
The number $d$ of the parameters $\{a\}$ is chosen
to be commensurate with current experimental constraints.
For this study we use $d=16.$
The detailed forms adopted for the initial functions
$\{f(x,Q_{0};\{a\})\}$ are not of particular concern
in this study, since we shall be emphasizing
results obtained by ranging over the full
parameter space.\footnote{%
In other words, for this paper, the PDF parameters $\{a\}$
play mostly the role of ``internal variables''.
In contrast, they occupy the center stage in
the companion paper \cite{Hesse}.}
The explicit formulas are given in Appendix \ref{sec:AppPdfs}
(where relevant PDFs from the results of our study are presented).
The $T_{I}(\{a\})$ are calculated as convolution integrals of the
relevant NLO QCD matrix elements and the universal parton distributions
$\{f(x,Q;\{a\})\}$ for all $Q$. The latter are obtained from the initial
functions $\{f(x,Q_{0};\{a\})\}$ by NLO QCD evolution.

The \emph{global analysis} consists of a systematic way to determine
the best values for the $\{a\},$ and the associated uncertainties,
by fitting $\{T(a)\}$ to $\{D\}$.
Because of the wide range of experimental and theoretical sources
of uncertainty mentioned in the Introduction, there are a variety
of strategies to deal with the complex issues
involved\ \cite{ZomerDis,Alekhin,GieleEtal,Botje,RunII}.
In the next two sections, the primary
tool we employ is conventional $\chi ^{2}$ analysis.
The important task is to define an effective $\chi^{2}$ function,
called $\chi_{\mathrm{global}}^{2}(a) $, that conveniently combines
the theoretical and global experimental inputs,
as well as relevant physics considerations based on prior knowledge,
to give an overall measure of the goodness-of-fit for a given set
of PDF parameters.

Experience in global analysis of PDFs during the past two decades has
demonstrated that the PDFs obtained by the minimization of such a suitably
chosen $\chi _{\mathrm{global}}^{2}$ provide very useful up-to-date hadron
structure functions which, although not unique, are representative of good fits
between theory and experiments.
Now we must quantify the uncertainties of the PDFs and their
predictions; {\it i.e.,} we must expand the scope of the work
from merely identifying typical solutions to systematically
mapping the PDF parameter space in the neighborhood around
the minimum of $\chi^2$.

The simplest possible choice for the $\chi ^{2}$ function would be
\begin{equation}
\chi^{2}(a)=\sum_{I=1}^{N}
\frac{\left[D_{I}-T_{I}(a)\right]^{2}}{\sigma
_{I}^{2}}  \label{eq:Ch2generic}
\end{equation}
where $\sigma _{I}$ is the error associated with data point $I$.
Through $T_{I}(a)$, $\chi^{2}(a)$ is a function of
the theory parameters $\{a\}$.
Minimization of $\chi^{2}(a)$ would identify
parameter values for which the theory fits the data.
However, the simple form (\ref{eq:Ch2generic}) is appropriate only
for the ideal case of a uniform data set with uncorrelated errors.
For data used in the global analysis, most experiments combine various
systematic errors into one effective error for each data point,
along with the statistical error.
Then, in addition, the fully correlated normalization error of
the experiment is usually specified separately.
For this reason,
it is natural to adopt the following definition for the effective $\chi^2$
(as done in previous CTEQ analyses):
\begin{eqnarray}
\chi _{\mathrm{global}}^{2}(a) &=&
\sum_{n} w_{n} \chi _{n}^{2}(a)\qquad (n\;%
\mbox{labels the different experiments})
\label{eq:Chi2global}
\\
\chi _{n}^{2}(a) &=&\left(\frac{1-{\cal N}_{n}}{\sigma _{n}^{N}}\right)^{2}
+\sum_{I}\left( \frac{{\cal N}_{n}D_{nI}-T_{nI}(a)}{\sigma _{nI}^{D}}
\right)^{2}
\label{eq:Chi2n}
\end{eqnarray}
For the $n^{\mathrm{th}}$ experiment, $D_{nI}$, $\sigma _{nI}^{D}$, and $%
T_{nI}(a)$ denote the data value, measurement uncertainty (statistical and
systematic combined), and theoretical
value (dependent on $\{a\}$) for the $I^{\mathrm{th}}$ data point; $\sigma
_{n}^{N}$ is the experimental normalization uncertainty;
${\cal N}_{n}$ is an overall normalization factor (with default
value $1$) for the data of experiment $n$.
The factor $w_{n}$ is a possible weighting factor (with default value $1$)
which may be necessary to take into account prior knowledge based on
physics considerations or other information.
The {\it a priori} choices represented by the $w_{n}$ values are
present, explicitly or implicitly, in any data analysis.
For instance, data inclusion or omission (choices which vary for
different global analysis efforts) represent extreme cases,
assigning either $100\%$ or \ $0\%$ weight to each available
experimental data set.
Similarly, choices of various elements of the
analysis procedure itself represent subjective input.
Subjectivity of this kind also enters into the analysis
of systematic errors in experiments.

The function $\chi_{\mathrm{global}}^{2}(a)$ allows the inclusion of all
experimental constraints in a uniform manner while allowing flexibility for
incorporating other relevant physics input.
We will make use of this function to explore the neighborhood
of the best fit, and to generate sample PDFs pertinent
to the uncertainty of the prediction of a specific physical
variable of interest.
However, the numerical value of this effective $\chi^{2}$
function should not be given an \textit{a priori} statistical
interpretation, because correlations between measurement errors,
and correlated theoretical errors, are not included in its definition.
In particular, the likelihood of a candidate PDF set $\{a\}$ cannot
be determined by the value of the increase
$\Delta\chi_{\mathrm{global}}^{2}(a)$
above the minimum.\footnote{%
The often quoted theorem of Gaussian error analysis, that an increase of
$\chi^{2}$ by 1 unit in a constrained fit to data corresponds to 1 standard
deviation of the constrained variable, is true only in the absence of
correlations.
When existing correlations are left out, the relevant size of
$\Delta\chi^2$ can be much larger than $1$.
Appendix \ref{sec:AppDelChi} discusses this point in some detail.}
Instead, the evaluation of likelihoods and estimation of global
uncertainty will be carried out in a separate step in
Sec.\,\ref{sec:quantifying}, after sets of
optimal sample PDFs for the physical variable of interest
have been obtained.

\section{The Lagrange Multiplier Method}
\label{sec:Lagrange}

The Lagrange Multiplier method is an extension of the $\chi^{2}$
minimization procedure, that relates
the range of variation of a physical observable $X$ dependent upon
the PDFs, to the variation of the function $\chi_{\mathrm{global}}^{2}(a)$
that is used to judge the goodness of fit
of the PDFs to the experimental data and PQCD.

\subsection{The Method}%
\label{sec:Method}

The method has been introduced in \cite{RunII,pdfuc0}.
The starting point is to perform a
global analysis as described in Sec.~\ref{sec:Global},
by minimizing the function $\chi_{\mathrm{global}}^{2}(a)$
defined by Eq.~(\ref{eq:Chi2global}), thus
generating a set of PDFs that represents the best estimate
consistent with current experiment and theory.
We call this set the ``standard set''\footnote{%
This standard set is very similar to the published
CTEQ5M1 set \cite{cteq5}.}, denoted $S_{0}$.
The parameter values that characterize this set will be
denoted by $\{a^{(0)}\}\equiv \{a^{(0)}_{i};\,i=1,\dots,d\}$;
and the absolute minimum of $\chi_{\mathrm{global}}^{2}$
will be denoted by $\chi_{0}^{2}$.
Now, let $X$ be a
particular physical quantity of interest.
It depends on the PDFs, $X=X(a)$, and
the best estimate (or \thinspace prediction)
of $X$ is $X_{0}=X(a^{(0)})$.
We will assess the {\em uncertainty} of this predicted value
by a two-step analysis.
First, we use the Lagrange Multiplier method to determine how
the minimum of $\chi _{\mathrm{global}}^{2}(a)$ increases,
\textit{i.e.,} how the quality of the fit to the
global data set decreases, as $X$ deviates from the best estimate $X_{0}$.
Second, in Section 4, we analyze the appropriate tolerance of
$\chi_{\mathrm{global}}^{2}$.

As explained in \cite{RunII,pdfuc0}, the first step is taken
by introducing a Lagrange multiplier variable $\lambda $,
and minimizing the function
\begin{equation}
\Psi (\lambda ,a)=\chi _{\mathrm{global}}^{2}(a)+\lambda X(a)
\label{eq:LMF}
\end{equation}
with respect to the original $d$ parameters $\{a\}$ for fixed
values of $\lambda$.
In practice we minimize $\Psi(\lambda,a)$ for many values
of the Lagrange multiplier:
$\lambda_{1},\lambda_{2},\dots, \lambda_{M}$.
For each specific value $\lambda_{\alpha}$, the minimum of
$\Psi(\lambda_{\alpha},a)$
yields a set of parameters $\{a_{\mathrm{min}}(\lambda_{\alpha})\}$,
for which we evaluate the observable $X$ and the related
$\chi^{2}_{\rm global}$.
We use the shorthand
$(X_{\alpha},\chi^{2}_{{\rm global},\alpha})$ for this pair.
$\chi^{2}_{{\rm global},\alpha}$ represents the lowest achievable
$\chi^{2}_{{\rm global}}$, for the global data, for which $X$ has
the value $X_{\alpha}$, taking into account all possible PDFs
in the {\em full $d$-dimensional parameter space} of points $\{a\}$.
In other words, the result $\{a_{\mathrm{min}}(\lambda_{\alpha})\}$
is a {\em constrained fit}---with $X$ constrained to be $X_{\alpha}$.
We can equivalently say that $X_{\alpha}$ is an extremum of $X$
if $\chi^{2}_{\rm global}$ is constrained to be
$\chi^{2}_{{\rm global},\alpha}$.
We denote the resulting set of PDFs by $S_{\alpha}$.

We repeat the calculation for many values of $\lambda$,
following the chain
\[
\lambda_{\alpha}  \longrightarrow  \mathrm{min}
\left[\Psi(\lambda_{\alpha},a)\right]
\longrightarrow a_{\rm min}(\lambda_{\alpha})
\longrightarrow X_{\alpha} {\rm ~and~}
\chi^{2}_{{\rm global},\alpha}
\]
for $\alpha=1, 2, 3,\dots,M$.
The result is a parametric relationship between $X$
and $\chi^{2}_{\rm global}$, through $\lambda$.
We call this function $\chi^{2}_{\rm global}(X)$; so $\chi^{2}_{\rm
global}(X_{\alpha})=\chi^{2}_{{\rm global},\alpha}$ is the minimum of
$\chi^{2}_{\rm global}(a)$ when $X$ is constrained to be $X_{\alpha}$.
The absolute minimum of $\chi^{2}_{\rm global}$, which we denote
$\chi^{2}_{0}$, is the minimum of $\Psi(\lambda=0,a)$, occurring
at $\{a\}=\{a^{(0)}\}$.
Thus the procedure generates a set of optimized sample PDFs along the
curve of maximum variation of the physical variable $X$ in the
$d$-dimensional PDF parameter space (with $d=16$ in our case).
These PDF sets $\{S_{\alpha}\}$ are exactly what is needed to assess
the range of variation of $X$ allowed by the data.
In other words, the Lagrange Multiplier method provides
optimal PDFs tailored to the physics problem at hand,
in contrast to an alternative method \cite{GieleEtal}
that generates a large sample of PDFs by the Monte Carlo method.
The underlying ideas of these two complementary approaches
are illustrated in the plot on the left side of
Fig.~\ref{fig:pedagogy}.
\figPedagogy

$\chi_{\mathrm{global}}^{2}(X)$ is the lowest achievable
value of $\chi_{\mathrm{global}}^{2}(a)$ for the value $X$ of the
observable, where $\chi_{\mathrm{global}}^{2}(a)$ represents our
measure of the goodness-of-fit to the global data.
Therefore the allowed range of $X$,
say from $X_{0}-\Delta{X}$ to $X_{0}+\Delta{X}$,
corresponding to a chosen tolerance of the goodness of fit
$\Delta\chi_{\mathrm{global}}^{2}
=\chi_{\mathrm{global}}^{2}-\chi_{0}^{2}$,
can be determined by examining
a graph of $\chi _{\mathrm{global}}^{2}$ versus $X$,
as illustrated in the plot on the right side of Fig.\ \ref{fig:pedagogy}.
This method for calculating $\Delta{X}$ may be more robust and reliable
than the traditional error propagation because it does not approximate
$X(a)$ and $\chi_{\mathrm{global}}^{2}(a)$ by linear and
quadratic dependence on $\{a\}$, respectively,
around the minimum.

Although the parameters $\{a\}$ do not appear explicitly in this
analysis, the results do depend, in principle, on the choice of
parameter space (including the dimension, $d$) in which the
minimization takes place.
In practice, if the degrees of freedom represented
by the parametrization are chosen to match the constraining power of
the global data sets used, which must be true for a sensible global
analysis, the results are quite stable with respect to changes
in the parametrization choices.
The sensitivity to these choices is tested,
as part of the continuing effort to improve the global analysis.

The discussion so far has left open this question:
What is the appropriate tolerance $\Delta\chi_{\mathrm{global}}^{2}$
to define the ``error'' of the prediction $X_{0}$?
This question will be addressed in Sec.~\ref{sec:quantifying}.

Our method can obviously be generalized to study the uncertainties of a
collection of physical observables $(X_{1},\,X_{2},\dots,X_{s})$ by
introducing a separate Lagrange multiplier for each observable.
Although the principle stays the same, the amount of computational
work increases dramatically with each additional observable.

\subsection{A case study: the ${W}$ cross section}

\label{sec:wXsec}

In this subsection we examine the cross section $\sigma_{W}$
for inclusive $W^{\pm}$ production at the Tevatron
($p\overline{p}$ collisions at $\sqrt{s}=1.8$\,TeV)
to illustrate the method and to lay the ground work for the
quantitative study of uncertainties to be given in Sec.\,4.
Other examples will be described in Sec.~\ref{sec:MoreExamples}.
Preliminary results of this section have been reported previously
\cite{RunII,pdfuc0}.

Until recently the only method for assessing the uncertainty of
$\sigma_{W}$ due to PDFs has been to compare the calculated values
obtained from a number of different PDFs, as illustrated in
Fig.~\ref{fig:WprodPDFcomparison}, in which the plots are taken
from existing literature.\footnote{%
These plots show the product of $\sigma_{W}$ times a leptonic
branching ratio, which is what is measured experimentally.
The branching ratio $B$ has some experimental error.
For studying the uncertainties of $\sigma_{W}$, we will
focus on $\sigma_{W}$ itself in the rest of the paper.}
The PDFs used in these comparisons are either the
``best fits'' from different global analysis groups \cite{cteq5, MRST}
(hence are not pertinent to uncertainty studies) or are chosen by some
simple intuitive criteria \cite{MRST2}.
The meaning and reliability of the resulting range of $%
\sigma _{W}$ are not at all clear. Furthermore, these results do not provide
any leads on how the uncertainties can be improved in the future. The Lagrange
Multiplier technique provides a systematic method
to address and remedy both of these problems.%
\figWprodPDFcomparison

Let the physical quantity $X$ of the last subsection be the cross section
$\sigma_{W}$ for $W^{\pm}$ production at the Tevatron. Applying the Lagrange
method, we obtain the constrained minimum of $\chi _{\mathrm{global}}^{2}$ as a
function of $\sigma _{W}$, shown as solid points in Fig.~\ref{fig:Wprod}. The
best estimate value, {\it i.e.,} the prediction for the standard set $S_0$,
is $\sigma_{W0}=21.75$\,nb.%
\figCsqvsWTev%
The curve is a polynomial fit to the points
to provide a smooth representation of the continuous
function $\chi^{2}_{\rm global}(X)$.
We see that all the sample PDF sets obtained by this method lie
on a smooth quasi-parabolic curve with the best-fit value at the
minimum.

As discussed earlier (in Fig.\ \ref{fig:pedagogy}) points on the curve
represent our sample of optimal PDFs relevant to the determination
of the uncertainty of $\sigma_{W}$.
To quantify this uncertainty, we need to reach beyond the effective
$\chi_{\rm global}$ function, and establish the confidence levels for
these ``alternative hypotheses'' with respect to the experimental data
sets used in the global analysis.


\section{Quantifying the Uncertainty}

\label{sec:quantifying}

Consider a series of sample PDF sets along the curve
$\chi _{\mathrm{global}}^{2}(X)$ of Fig.~\ref{fig:Wprod}
denoted by $\{S_{\alpha};\alpha=0,1,\dots,M\}$
where $S_{0}$ is the standard set.
These represent ``alternative hypotheses'' for the true PDFs,
and we wish to evaluate the likelihoods associated with
these alternatives.
To do so, we go back to the individual experiments and,
in each case, perform as detailed a statistical analysis
as is permitted with available information from that experiment.
After we have obtained meaningful estimates of the ``errors'' of
these candidate PDFs with respect to the individual experiments,
we shall try to combine this information into a global
uncertainty measure in the form of $\Delta X$
and $\Delta \chi_{\rm global}^{2}$.

\tblDatSet
The experimental data sets included in our global analysis are listed in
Table \ref{tbl:DatSet}.  For some of these experiments,
information on correlated systematic errors is available (albeit usually in
unpublished form).
For these, statistical inference should be drawn from a more accurate
$\chi_{n}^{2}$ function
than the simple formula Eq.~(\ref{eq:Chi2n}) used for the global fit.
In particular, if $\sigma _{nI}$ is the uncorrelated error
and $\{\beta _{kI};\,k=1,2,\dots,K\}$ are the coefficients of $K$ distinct
correlated errors associated with the data point $I$, then an
appropriate formula for the $\chi_{n}^{2}$ function is
\begin{equation}
\chi_{n}^{2}=\sum_{I}\frac{(D_{nI}-T_{nI})^{2}}{\sigma_{nI}^{2}}%
-\sum_{k=1}^{K}\sum_{k^{\prime }=1}^{K}B_{k}\left( A^{-1}\right)
_{kk^{\prime }}B_{k^{\prime }}  \label{eq:chi_corr}
\end{equation}
where $B_{k}$ is a vector, and $A_{kk^{\prime }}$ a matrix, in $K$
dimensions:
\begin{equation}
B_{k}=\sum_{I}\beta _{kI}(D_{nI}-T_{nI})/\sigma_{nI}^{2}
\;\;;\;\;A_{kk^{\prime }}=\delta _{kk^{\prime }}
+\sum_{I}\beta_{kI}\beta _{k^{\prime }I}/\sigma _{nI}^{2}.
\label{eq:BandA}
\end{equation}
(The sum over $I$ here includes only the data from experiment $n$.)
Traditionally, $\chi^{2}_{n}$ is written in other ways,
{\it e.g.,} in terms of the inverse of the ($N\times N$)
variance matrix.
For experiments with many data points, the inversion of such
large matrices may lead to numerical instabilities, in addition
to being time-consuming.
Our formula (\ref{eq:chi_corr}) has a significant advantage in that
all the systematic errors are first combined (``analytically'') in
the definitions of $B_{k}$ and $A_{kk'}$.
Equation (\ref{eq:chi_corr}) requires only the inverse of the
much smaller ($K\times K$) matrix $A_{kk'}$.
($K$ is the number of distinct systematic errors.)
The derivation of these formulas is given in
Appendix \ref{sec:AppCorSys}.
Equation~(\ref{eq:chi_corr}) reduces to the minimum of
$\chi_{n}^{2}$ in Eq.~(\ref{eq:Chi2n}) with respect to
${\cal N}_{n}$ if the only correlated error is the overall
normalization error for the entire data set;
in that case $\beta_{I}=-\sigma^{N}_{n}D_{nI}$.

By using Eq.~(\ref{eq:chi_corr}), or
Eq.~(\ref{eq:Chi2n}) for cases where the correlations of
systematic errors are unavailable,
we obtain the best estimate on the range of
uncertainty permitted by available information
on each individual experiment.
We should note that the experimental data sets are continuously evolving.
Some data sets in Table 1 will soon be updated (Zeus, H1) or replaced
(CCFR).\footnote{%
{\it Cf.}\ Talks presented by these collaborations at DIS2000 \emph{Workshop on
Deep Inelastic Scattering and Related Topics}, Liverpool, England, April
2000.}
In addition, most information on correlated systematic errors
is either unpublished or preliminary.
The results presented in the following analysis should therefore
be considered more as a demonstration of principle---as the first
application of our proposed method---rather than the final word
on the PDF uncertainty of the $W$ cross section.

\subsection{Uncertainty with respect to individual experiments}

As an example, we begin by comparing the $\{S_{\alpha}\}$
series for $\sigma _{W}$
at the Tevatron to the H1 data set \cite{NewH1}.
Results on correlated systematic errors are available
for this data set,\footnote{%
These systematic errors are unpublished results,
but are made available to the public on the H1 Web page.
For convenience, we have approximated each of the pair
of 4 non-symmetrical errors by a single symmetric error.
The size of the resulting error on $\sigma _{W}$ inferred
from this evaluation is not affected by that approximation.}
and are incorporated in the calculation using Eq.~(\ref{eq:chi_corr}).
The number of data points in this set is $N_{H1}=172$.
The calculated values of $\chi _{H1}^{2}/N_{H1}$ are plotted
against $\sigma_{W}$ in Fig.~\ref{fig:H1Tev}. The curve is a smooth
interpolation of the points. The value of $\chi _{H1}^{2}/N_{H1}$ for the
standard set $S_{0}$ (indicated by a short arrow on the plot) is $0.975$;
and it is $0.970$ at the minimum of the curve.
These values are quite normal for data with
accurately determined measurement errors.
We can therefore apply standard statistics to calculate the $90$\%
confidence level on $\chi ^{2}/N$ for $N=172$. The result is shown as the
dashed horizontal line in Fig.~\ref{fig:H1Tev}.\figHoneTev

We have similarly calculated $\chi_{n}^{2}/N_{n}$ including information
on the correlations of systematic errors for the BCDMSp data set.
The results are similar to the H1 results, except that the absolute
values are all larger than 1.12, a large value for $N=168$ data points.
This is a familiar problem in data analysis, and it is
encountered in several other data sets in this global analysis
({\it cf.}\ below).
The $\chi_{n}^{2}/N_{n}$ calculation including correlations of the errors
is also done for the D0 and CDF jet cross sections.\footnote{%
The measurement errors of the jet cross sections
are dominated by systematic errors, so the error correlation
matrices are used for $\chi^{2}_{n}$ of these experiments
even in $\chi^{2}_{\mathrm{global}}$.}
For those experiments that have only provided (effective) uncorrelated errors,
we must rely on Eq.~(\ref{eq:Chi2n}) for our error calculation, since that
represents the best information available.

In order to obtain usable likelihood estimates from all the data sets, one must
address the problem mentioned in the previous paragraph:
Even in a ``best fit'', the values of\ $\chi ^{2}$ per data
point, $\chi_{n}^{2}/N_{n},$ for individual experiments vary considerably
among the established experiments (labeled by $n$).
Specifically, $\chi_{n}^{2}/N_{n}$ ranges from $1.5-1.7$
(for ZEUS and CDFjet) on the high end to $0.5-0.7$
(for some D-Y experiments) on the low end in all good fits.
Considering the fact that some of these data sets contain close to $200 $
points, the range of variation is huge from the viewpoint
of normal statistics:
Experiments with $\chi _{n}^{2}$ $/N_{n}$ deviating from $1.0$
by a few times $\sqrt{2/N_{n}}$ in either direction would have
to be ruled out as extremely unlikely \cite{Kosower98}.

The reasons for $\chi _{n}^{2}/N_{n}$ to deviate from $1.0$ in real
experiments are complex, and vary among experiments.
They are, almost by definition, not understood, since
otherwise the errors would have been corrected and
the resulting $\chi^2$ would become consistent with
the expected statistical value.
Under these circumstances, a commonly adopted pragmatic
approach is to focus on the relative $\chi^{2}$ values with
respect to some appropriate reference $\chi^{2}$.\footnote{%
The alternative is to take the \emph{absolute} values of
$\chi_{n}^{2}$ seriously, and hence only work with alternative
hypotheses and experiments that are both self-consistent
({\it i.e.,} have $|\chi_{n}^{2}/N_{n}-1|\lesssim \sqrt{2/N_{n}}$)
and mutually compatible in the strict statistical sense
({\it i.e.,} have overlapping likelihood functions).
Since few of the precision DIS experiments are compatible in this sense,
one must then abandon global analysis, and work instead with several
distinct (and mutually exclusive) analyses based on different
experiments.}
Accordingly, in the context of performing {\em global} QCD analysis,
we adopt the following procedure.
For each experiment (labeled by $n$):

\begin{Simlis}[]{0.5em}
\item (i) Let $\chi^{2}_{n,0}$ denote the value of $\chi^{2}_{n}$
for the standard set $S_{0}$.
We assume $S_{0}$ is a viable reference set.
Because $\chi^{2}_{n,0}$ may be far from a likely value
for random errors, we {\em rescale} the values of
$\chi^{2}_{n,\alpha}$ (for $\alpha=0, 1, 2,\dots, M$)
by a factor $C_{n0}$, calling the result
$\overline{\chi}^{2}_{n,\alpha}$
\begin{equation}\label{eq:rescale}
\chi^{2}_{n,\alpha} \longrightarrow
\overline{\chi}^{2}_{n,\alpha} \equiv
C_{n0}\,\chi^{2}_{n,\alpha}.
\end{equation}
The constant $C_{n0}$ is chosen such that, for the standard set,
$\overline{\chi}^{2}_{n,0}$ assumes the most probable value for
a chi-squared variable:
$\overline{\chi}^{2}_{n,0} \! = \! \xi_{50} \equiv$ the 50$^{th}$ percentile
of the chi-squared distribution $P(\chi^{2},N_{n})$
with $N_{n}$ degrees of freedom, defined by
\begin{equation}
\int_{0}^{\xi_{50}} P(\chi^{2},N_{n}) d\chi^{2} = 0.50.
\end{equation}
(If $N_{n}$ is large then $\xi_{50} \! \approx \! N_{n}$.)
The rescaling constant $C_{n0}$ is thus $\xi_{50}/\chi^{2}_{n,0}$.
For random errors the probability that $\chi^{2} \! < \! \xi_{50}$
(or $> \! \xi_{50}$) is 50\%.
For those experiments whose $\chi^{2}_{n,0}$ deviates significantly
from $\xi_{50}$,
this rescaling procedure is meant to
provide a simple (but crude) way to correct for the unknown correlations
or unusual fluctuations.

\item (ii) We then examine the values of
$\overline{\chi}^{2}_{n,\alpha}$ for the alternative sets
$S_{\alpha}$ with $\alpha=1, 2, \dots, M$,
using $\overline{\chi}^{2}_{n,\alpha}-\overline{\chi}^{2}_{n,0}$
to compute the statistical likelihood of the
alternative hypothesis $S_{\alpha}$ with respect to the
data set $n$, based on the chi-squared distribution
with $N_{n}$ data points.
\end{Simlis}
This procedure does not affect the results presented earlier for the H1
experiment, since $\chi^{2}_{n,0}/N_{n}$ is already very close to $1$
for that experiment.

Before presenting the results of the likelihood calculation, it is
interesting to examine, in Fig.~\ref{fig:allTev},
the differences
$\Delta \chi_{n,\alpha}^{2}=\chi_{n,\alpha}^{2}-\chi_{n,0}^{2}$
(before rescaling) versus $\sigma_{W}$ for the 15 data sets.
(N.B.\ The vertical scales of the various plots are not the same,
due to the large variations in the value of $\Delta \chi^{2}_{n,\alpha}$
for different experiments.)
The ordering of the experiments in Fig.\ \ref{fig:allTev}
is the same as in Table \ref{tbl:DatSet}, with experiments ordered by
process (DIS, DY, $W$ and jet production).
It is clear from these graphs that the DIS experiments place
the strongest constraints on $\sigma _{W}$, because they have
the largest $\Delta \chi_{n}^{2}$ for the same $\Delta \sigma _{W}$.
This is to be expected since quark-antiquark annihilation makes the
dominant contribution to $\sigma_{W}$.
We also observe that most experiments place some constraint on
$\sigma_{W}$ on both sides, but a few bound it on one side only.
Globally, as shown in Fig.~\ref{fig:Wprod}, the combined constraints
give rise to a classic parabolic behavior for
$\chi^{2}_{\rm global}(\sigma _{W})$.
\figallTev

To estimate the statistical significance of the
individual $\chi_{n}^{2}$ increases,
we assume that the rescaled variable
$\overline{\chi}^{2}_{n}$ obeys a chi-squared distribution
$P(\chi^{2},N_{n})$ for $N_{n}$ data points.
Thereby, we estimate the
value of ${\overline{\chi}}_{n}^{2}$ that corresponds to
the 90\% confidence level (CL) uncertainty for $\sigma _{W}$
(with respect to experiment $n$) from the formula
$\overline{\chi}_{n}^{2}=\xi_{90}$, where $\xi_{90}$ is
the 90$^{th}$ percentile defined by
\begin{equation}\label{eq:xi90}
\int_{0}^{\xi_{90}} P(\chi^{2},N_{n}) d\chi^{2} = 0.90.
\end{equation}
For example,
Fig.~\ref{fig:chsqdist} shows the chi-squared distribution
$P(\chi^{2},N_{n})$ for $N_{n}=172$, the number of data
points in the H1 data set.
The 50$^{th}$ and 90$^{th}$ percentiles are indicated.
We choose a conservative 90\% CL because
there are other theoretical and phenomenological uncertainties not taken
into account by this analysis.
\figchsqdist

To summarize our procedure, an alternative PDF set
$S_{\alpha}$ lies within the 90\% CL for experiment $n$ if it has
$\overline{\chi}^{2}_{n,\alpha}<\xi_{90}$; that is, if
\begin{equation}
\frac{\chi^{2}_{n,\alpha}}{\chi^{2}_{n,0}}
< \frac{\xi_{90}}{\xi_{50}}.
\end{equation}
We judge the likelihood of $S_{\alpha}$ from the {\em ratio}
of $\chi^{2}_{n,\alpha}$ to the reference value $\chi^{2}_{n,0}$,
rather than from the absolute magnitude.
The horizontal lines in Fig.~\ref{fig:allTev} correspond to the
values of $\Delta\chi_{n}^{2}$ obtained in this way.
Finally, from the intercepts of the line with the interpolating
curve in each plot in Fig.~\ref{fig:allTev}, we obtain an
estimated uncertainty range of $\sigma _{W}$ from each individual
experiment.
The results are presented collectively in Fig.~\ref{fig:rangesTev},
where, for each experiment, the point ($\bullet $) is the value of
$\sigma _{W}$ for which $\chi _{n}^{2}$ is minimum, and the error
bar extends across the 90\% CL based on that data set.
\figrangesTev%

The uncertainty ranges shown in Fig.\ \ref{fig:rangesTev}
with respect to individual experiments represent the
most definitive results of our study, in the sense that the input and the
assumptions can be stated clearly and the analysis is quantitative within
the stated framework.
It is natural to proceed further and estimate a global measure of
$\Delta\sigma_{W}$ and the corresponding $\Delta\chi_\mathrm{global}^{2}$.
This last step is, however, less well-defined and
requires some subjective judgement.

\subsection{The Global Uncertainty}
\label{subsec:globaluc}

It should be emphasized that the ranges shown by the error bars in Fig.~\ref
{fig:rangesTev}\ are not errors determined independently by each experiment;
rather they represent the ranges allowed by the experiments for alternative
{\em global} fits $\{S_{\alpha}\}$. For this reason, and others related to the
rescaling of $\chi ^{2}$ mentioned earlier as well as approximations inherent
in many of the original published
errors,\footnote{%
For instance, the single uncorrelated systematic error associated with each
data point, which is the only systematic error given for most experimental data
sets, is clearly only an ``effective uncorrelated error'' which qualitatively
represents the effects of the many sources of systematic error, some of which
are really correlated.} it is not obvious how to {\em combine} these errors.
We refer to the ranges in Fig.\ \ref{fig:rangesTev} by the generic term
{\em local} ({\it i.e.,} single-experiment) {\em uncertainties}.
On a qualitative level, Fig.~\ref{fig:rangesTev} exhibits the same features
seen earlier in Fig.~\ref {fig:allTev}: (i) the quark dominated DIS
experiments give the smallest error bars;
and (ii) a few experiments only set bounds on one side,
while the rest limit the range in both directions.
In addition, Fig.~\ref{fig:rangesTev} gives us an overall view which
clearly shows that $\sigma_{W}$ is well constrained in the global
analysis, and the experimental bounds are consistent with each other.

The important question is how to provide a sensible measure of the overall
uncertainty in view of the complexity of the problem already described. The
situation here is not unlike the problem of assigning an overall systematic
error to an experimental measurement. Figure \ref{fig:rangesTev} shows a set of
90\% CL ranges for $\sigma_{W}$ from different sources, but these ranges are
highly correlated, because the alternative hypotheses being tested come from
global fits. The final uncertainty
must be a reasonable intersection of these ranges.

We will state an algorithm for obtaining the final uncertainty
measure of $\sigma_{W}$ based on Fig.\ \ref{fig:rangesTev}.
The same algorithm can be applied in the future for predictions
of other observables.
It has two parts:
(1) Determine the central value using all the experiments;
that is the solid line in Fig.\ \ref{fig:rangesTev}.
(2) Then take the {\em intersection} of the error
ranges as the combined uncertainty.
But in calculating the intersection, experiments below the
mean are used only for setting the lower bound, and experiments
above the mean are used only for setting the upper bound.
With this algorithm, experiments that permit a large range
of $\sigma_{W}$,
{\it i.e.,} that depend on aspects of the PDFs that are not
sensitive to the value of $\sigma_{W}$,
will not affect the final uncertainty measure (as they should not).
According to this algorithm,
the result for the uncertainty of $\sigma_{W}$ is
$20.9\,{\rm nb} \! < \! \sigma_{W} < 22.6\,{\rm nb}$.
These bounds are approximately $\pm{4}$\% deviations
from the prediction (21.75\,nb) and so we quote
a $\pm{4}$\% uncertainty in $\sigma_{W}$ due
to PDFs.

Now we may determine the increase in $\chi^{2}_{\rm global}$
that corresponds to our estimated uncertainty
$\Delta\sigma_{W}$ in the $\sigma_{W}$ prediction.
Referring to Fig.\ \ref{fig:Wprod},
a deviation of $\sigma_{W}$ by $\pm{4}$\% from the minimum
corresponds to an increase
$\Delta\chi^{2}_{\rm global} \! \approx \! 180$.
That is, $\Delta\chi^{2}_{\rm global}$ in
Fig.\ \ref{fig:pedagogy} is 180.
In other words, along the direction of maximum variation
of $\sigma_{W}$ a PDF set with
$\Delta\chi^{2}_{\rm global} \gtrsim 180$ is found
to violate some experimental constraints by this analysis.

\subsection{Comments}

We should point out that the above uncertainty estimate,
$\Delta\sigma_{W}/\sigma_{W} \sim 4$\%,
represents only a lower bound on the true uncertainty,
since many other sources of error have not yet been
included in the analysis: theoretical ones such as
QCD higher order and resummation effects, power-law corrections,
and nuclear corrections.
These need to be taken into consideration in a
full investigation of the uncertainties, but
that goes beyond the scope of this paper.%
\footnote{%
Because there are these additional sources of uncertainty,
we have used 90\% CL's, rather than 68\% CL's,
to calculate the error.} %
We shall add only two remarks which are more directly related
to our analysis.

The first concerns a technical detail.
In the results reported so far, we have fixed the normalization
factors \{${\cal N}_{n}$\} in the definition of $\chi_{\rm global}^{2}$
(Eq.~(\ref{eq:Chi2global})) at their values determined in the
standard fit $S_{0}$.
If we let these factors float when we perform the
Lagrange Multiplier analysis, $\Delta \sigma _{W}$ will increase
noticeably compared to Fig.\ \ref{fig:Wprod} for the same
$\Delta \chi _{\rm global}^{2}$.
However, upon closer examination, this behavior can be easily
understood and it does not imply a real increase in the
uncertainty of $\sigma_{W}$.
The key observation is that the additional increase (or decrease)
in $\sigma_{W}$ is entirely due to a \emph{uniform} increase
(or decrease) of \{${\cal N}_{n}$\} for all the DIS experiments.
There is a simple reason for this: The values of the $q$ and $\bar{q}$
distributions in the relevant $x$ range (which determine the value of
$\sigma_{W}$) are approximately proportional to
\{${\cal N}_{n}$\}$_{DIS}$.
Although every experiment does have a normalization uncertainty,
the probability that the normalization factors of all the
\emph{independent} DIS experiments would shift in the \emph{same}
direction by the \emph{same} amount is certainly unlikely.
Hence we avoid this artificial effect by fixing \{${\cal N}_{n}$\}
at their ``best values'' for our study.
Allowing the factors \{${\cal N}_{n}$\} to vary {\em randomly}
(within the published experimental normalization uncertainties)
would not change our estimated value of $\Delta\sigma_{W}$
significantly.

The second remark concerns the choice of parametrization.
We have mentioned that even the robust Lagrange Multiplier method
depends in principle on the choice of the parton parameter space,
{\it i.e.,} on the choice of the functional forms used for the
nonperturbative PDFs at the low momentum scale $Q_{0}$.
To check how our answers depend on the choice of parametrization
in practice, we have done many similar calculations, using different
numbers of free parameters within the same functional form
({\it cf.}\ Appendix \ref{sec:AppPdfs}),
and using different functional forms for the factor
multiplying $x^{a}\left( 1-x\right)^{b}$.
We have not seen any dependence of the uncertainty estimates
on these changes.
Although more radical ways of parametrizing the nonperturbative
PDFs might affect the result more, there is no known example of
such a parametrization, which at the same time still provides
an equally good fit to the full data set.

\section{Further Examples}

\label{sec:MoreExamples}

\subsection{$W^{\pm}$ production at the LHC}

A study similar to the last section has been carried out for inclusive
$W^{\pm}$ production at the LHC.
Figure \ref{fig:WprodLHC} %
\figCsqvsWLHC%
shows $\chi_{\mathrm{global}}^{2}$ versus
$\sigma _{W}$ for the process $pp\rightarrow W^{\pm }X$
at $\sqrt{s}=14$\thinspace TeV, summed over the two final states.
The curve is a smooth interpolation of a series of PDF sets
$\{S_{\alpha}\}$ generated by the Lagrange Multiplier method.
The best estimate value of the LHC cross section is
$\sigma_{W}=189.7$\,nb.

Comparing Figs.\ \ref{fig:Wprod} and \ref{fig:WprodLHC},
one immediately notices that the uncertainty of $\sigma_{W}$(LHC)
is greater than that of $\sigma_{W}$(Tevatron) for the same
$\Delta\chi^{2}_{\mathrm{global}}$.
This indicates that because $W$ production in $pp$ collisions
at the LHC and $\bar{p}p$ collisions at the Tevatron involve
different mixtures of parton subprocesses as well as different
kinematic ranges,
the constraints imposed by current experiments included in the
global analysis are also different for the two cases.
Referring to the map of the $d$-dimensional PDF parameter
space on the left side of Fig.~\ref{fig:pedagogy},
we are generating sample PDFs along different directions
$L_{X}$ in the two cases.
Therefore it is not surprising that the rate of variation
of $\chi^{2}_{\mathrm{global}}$ is also different.

To demonstrate this point,
and to quantify the uncertainty on the LHC prediction,
we have carried out the same error analysis as in
Sec.\,\ref{sec:quantifying},
{\it i.e.,} comparing the alternative
PDFs to the individual experiments.
Figure \ref{fig:rangesLHC} %
\figrangesLHC%
gives the final overview of the
90\% CL ranges of $\sigma _{W}$ obtained from these
comparisons, analogous to Fig.\ \ref{fig:rangesTev}
for the Tevatron cross section.
There are some differences compared to the Tevatron case.
The LHC prediction is more tightly constrained by experiments
that are sensitive to PDFs at small $x$.
This makes sense, because $W$ production at the LHC
is not dominated by valence $q\bar{q}$ interactions.
We note in particular that the two inclusive jet production
experiments place significant constraints on $\sigma_{W}$
at the LHC.

We can combine the individual error bars in Fig.\ \ref{fig:rangesLHC}
according to the algorithm proposed in Sec.\ \ref{sec:quantifying}
to produce a global uncertainty
measure for $\sigma _{W}$ at the LHC.
The lower bound on $\sigma_{W}$(LHC) obtained by the intersections
of the individual ranges is $\sigma_{W}=175.3$\,nb;
the upper bound is $\sigma_{W}=204.6$\,nb.
These bounds, shown as the dashed lines in Fig.\ \ref{fig:rangesLHC},
correspond to $\pm{8}$\% deviations from the prediction (189.7\,nb).
The global uncertainty on $\sigma_{W}$(LHC) is thus significantly
larger than that on $\sigma_{W}$(Tevatron).
Reinforcing this conclusion is the fact that the scatter of the points
in Fig.\ \ref{fig:rangesLHC} is larger than in Fig.\ \ref{fig:rangesTev}.

We can again inspect the increase in $\chi^{2}_{\rm global}$
from the quoted range of alternative fits.
A $\pm{8}$\% deviation from the minimum, symmetrized for
simplicity, corresponds to the increase
$\Delta\chi^{2}_{\rm global}\approx 200$.
This number is similar to the increase in $\chi^{2}_{\rm global}$
for our estimated uncertainty of $\sigma_{W}$
at the Tevatron.

In the companion paper \cite{Hesse}, we make some process-independent
estimates of $\Delta \chi _{\mathrm{global}}^{2}$ based on completely
different considerations.
Those arguments also yield the same order-of-magnitude estimates
of $\Delta\chi_{\mathrm{global}}^{2}$ (in the range from 100 to 200)
for acceptable PDFs around the global minimum.
Since the effective $\chi_{\mathrm{global}}^{2}$, as a measure of
goodness-of-fit, does not have a normal statistical implication,
points on a constant $\chi_{\mathrm{global}}^{2}$ surface in the PDF
parameter space do not necessarily correspond to a constant likelihood.
Some variation with the direction in the multi-dimensional space is to
be expected.

\subsection{Uncertainties on $Z^{0}$ production}

\label{sec:Z}

We conclude this section by presenting results from applying the Lagrange
Multiplier method to $Z^{0}$ production at the Tevatron and the LHC.

Figure \ref{fig:Z0prod}a shows the minimum $\chi_{\mathrm{global}}^{2}$ as a
function of $\sigma _{Z}$ at the Tevatron.
The global prediction is $\sigma_{Z}=6.55$\,nb.
The experimental measurement by the D0 collaboration is
$\sigma_{Z}B=0.221\pm 0.003\pm 0.011$\thinspace nb;
the result from CDF (all data from Run I) is
$\sigma_{Z}B=0.250 \pm 0.004 \pm 0.010$\thinspace nb.
(Here $B$ is the branching ratio for $Z^{0}\rightarrow e\bar{e}$,
which is $(3.367\pm 0.005)\times 10^{-2}$.)
The comparison of the prediction to Tevatron data
is discussed below.
Analyzing the local $\chi^{2}_{n}$ in the manner of
Sec.\ \ref{sec:quantifying}, in order to quantify the uncertainty
of the prediction, we find that the uncertainty of
$\sigma_{Z}$(Tevatron) due to PDFs is $\pm{3}$\% of the prediction.
The corresponding increase in $\chi^{2}_{\rm global}$,
symmetrized for simplicity, is approximately 130.

\figZzeroprod

For the LHC process $pp\rightarrow Z^{0}X$, Fig.~\ref{fig:Z0prod}b shows
the minimum $\chi _{\mathrm{global}}^{2}$ as a function of $\sigma _{Z}$.
The dependence of $\chi^{2}_{\rm global}$ on $\sigma_{W/Z}$(LHC)
exhibits a behavior departing from quadratic over the full range
of $\sigma_{W/Z}$ under study.
This is evidence that the Lagrange multiplier method can go beyond
the traditional error matrix approach (which depends on the quadratic
approximation) in exploring the neighborhood of the minimum.

The global prediction is $\sigma_{Z}=58.0$\thinspace nb.
Analyzing the local $\chi^{2}_{n}$ as in the other cases,
we find that the uncertainty of $\sigma_{Z}$(LHC) due to PDFs
is approximately $\pm{10}$\%.
As in the case of $W^{\pm}$ production, the PDF uncertainty for
$Z^{0}$ production at the LHC is significantly larger than that
at the Tevatron.
Measurement of $W^{\pm}$ and $Z^{0}$ production at the
LHC will therefore provide significant information on PDFs.

\subsection{Comparison with existing data}

\label{sec:TevData}

\exptandthband
For $W$ and $Z$ production at the Tevatron, we can compare our
calculated cross sections $\sigma_{W}$ and $\sigma_{Z}$,
with their ranges of uncertainty $\pm{4}$\% and $\pm{3}$\% respectively,
to the measurements of CDF and D0 from Run I. \cite{FLehner}
The comparison is shown in Fig.\ \ref{fig:exptandthband}.
The two experiments do not measure $\sigma_{W}$ and $\sigma_{Z}$
{\it per se}, but rather $\sigma_{W}B_{W}$ and $\sigma_{Z}B_{Z}$
where $B_{W}$ is the branching ratio for $W^{-}\rightarrow e\bar{\nu}$
and  $B_{Z}$ is the branching ratio for $Z^{0}\rightarrow e\bar{e}$.
We have used the values $B_{W}=0.106$ and
$B_{Z}=0.0337$ for the calculations \cite{PartDatGr}.
The bands in Fig.\ \ref{fig:exptandthband} show the
ranges of $\sigma_{W}B_{W}$ and $\sigma_{Z}B_{Z}$
from our PDF uncertainty study (but no uncertainty included
from $B_{W}$ and $B_{Z}$).
The two measurements of $\sigma_{W}B_{W}$ are consistent
with the uncertainty range.
The two measurements of $\sigma_{Z}B_{Z}$ are not.

It should be noted that CDF and D0 use different normalizations
for their luminosity determinations.
The CDF collaboration bases its luminosity purely on its own
measurement of the inelastic $\bar{p}p$ cross section
\cite{FAbe,MAlbrow}, while D0 uses the world average
for this cross section.
Thus current luminosities quoted by CDF are $6.2$\% lower than
those quoted by D0.
Consequently, all CDF cross section measurements are
{\it ab initio} $6.2$\% higher than those of D0.
If the CDF/D0 measurements of $\sigma_{W}B_{W}$ and
$\sigma_{Z}B_{Z}$ are rescaled by $6.2$\% with respect
to each other, they are in excellent agreement.

Because of the uncertainty in the inelastic
$\bar{p}p$ cross section, it has been proposed to
normalize future Tevatron (and LHC)
physics cross sections to the measured $W$ cross section
(or rather $\sigma_{W}B_{W}$).
This makes the determination of the uncertainty of
$\sigma_{W}$ due to PDFs even more important.

\subsection{Comparison with previous uncertainty estimates}

It is interesting to contrast our results to existing estimates of
the uncertainties of $\sigma_{W}$ and $\sigma_{Z}$ at the Tevatron
and LHC colliders based on the traditional method of comparing
results obtained from somewhat {\it ad hoc} PDFs.
Some of these previous comparisons for
$\sigma_W$(Tevatron) between historical PDFs as well as various
trial up/down sets obtained by the CTEQ and MRST groups were
shown in Fig.~\ref{fig:WprodPDFcomparison}.
We will briefly comment on the results of
\cite{MRST2} in the context of this paper.

Reference \cite{MRST2} constructs an extended set of MRST PDFs,
of which the most important for $\sigma_{W}$ and $\sigma_{Z}$
are the standard set MRST99 and three pairs of up/down sets designated
\[
\{\alpha_{S}\uparrow, \alpha_{S}\downarrow\}, \; \{g\uparrow, g\downarrow\},
\; \{q\uparrow, q\downarrow\}
\]
in which some aspect of the parton distributions is either raised ($\uparrow$)
or lowered ($\downarrow$) by an amount that represents an educated guess of a
``standard deviation''. The predictions of $\sigma_{W}$ and $\sigma_{Z}$ are
then compared for these alternative PDF sets to get an idea of the uncertainty
due to PDFs.

In the case of the Tevatron processes, the deviations of $\sigma_{W}$
or $\sigma_{Z}$ from the value for MRST99 for sets
$\{\alpha_{S}\uparrow,\alpha_{S}\downarrow\},
 \; \{g\uparrow, g\downarrow\}, \; \{q\uparrow,q\downarrow\}$
were found to be $\pm{2}$\%, $\pm{1}$\%, $\pm{3}$\%
respectively.
From these results, the authors of \cite{MRST2} concluded that the
uncertainties of $\sigma_{W}$ and $\sigma_{Z}$ at the Tevatron are
no more than about $\pm{4}$\%, and mainly attributable to the
normalization uncertainty in the input $u$ and $d$ distributions.
This conclusion appears to be quite consistent with the results
of the previous sections based on exploring the variation of the
cross section over the entire PDF parameter space.
(This range of uncertainty also happens to coincide with what
one would get by comparing historical PDF sets, as shown in the
right plot of Fig.~\ref{fig:WprodPDFcomparison}.)

For the LHC, the MRST study found that the uncertainty of $\sigma_{W}$
and $\sigma_{Z}$ at the LHC is only slightly larger than at the Tevatron;
the uncertainty was estimated to be $\pm{5}$\%.
The largest observed variations came from the sets
$\alpha_{S}\uparrow$ and $\alpha_{S}\downarrow$, differing from the
standard prediction by $\pm{4}$--$5$\%.
This estimate is \emph{considerably lower} than the $\pm{8}$--$10$\%
result obtained by the detailed analysis of the previous sections.
We have verified that the PDF sets that give $\pm{8}$--$10$\%
deviations of $\sigma_{W}$(LHC) and $\sigma_{Z}$(LHC) from
the standard prediction (represented by the points at the outer
edges of the corresponding plots in Figs.~\ref{fig:WprodLHC}
and \ref{fig:Z0prod}) provide equally good or better fits to the
global data sets compared to the fits of
\{$\alpha_{S}\uparrow$,\, $\alpha_{S}\downarrow$\}.\footnote{%
The values of $\chi_{\mathrm{global}}^{2}$
for \{$\alpha_{S}\uparrow$,\, $\alpha_{S}\downarrow$\}
are \{$1531,\; 1356$\}
compared to $\sim 1400$ for the outermost LHC sets shown in
Figs.~\ref{fig:WprodLHC} and \ref{fig:Z0prod} and presented
in the next section.}$^,$\footnote{%
The global fits used at first in our exploration of PDF uncertainty
were conducted with fixed $\alpha_s$.
To make sure that this restriction does not result in an
underestimate of the uncertainties of $\sigma_{W}$ and $\sigma_{Z}$,
we have examined the effect of freeing $\alpha_s$ in the analysis
(but imposing the
known constraints from the world average of $\alpha_s$).
The results on the size of the uncertainties are not changed noticeably.
This is because the full variations in the PDFs (particularly the gluon)
allowed in the Lagrange approach can absorb the added degree of freedom.}
Thus, it is clear that the Lagrange Multiplier method can generate
{\em optimal} PDFs, {\it i.e.,} having the {\em largest} excursion
of the variable $X$ of interest, which are difficult to discover by
{\it ad hoc} trial-and-error methods used in the past.

\section{PDF Sets for exploring W and Z Physics}

\label{sec:UpDownPdf}

The parton distribution sets used in the above calculations are useful
for exploring some aspects of $W$ and $Z$ physics at the Tevatron and
LHC, since they provide much more reliable estimates on the PDF
uncertainties than existing ones in the literature, which are
not designed to probe the full range of possibilities in the
parton parameter space.
With this in mind, we present in this section some representative
PDF sets for applications to the rate of $W$ and $Z$ production
at the Tevatron and LHC.
These PDFs are relevant to the total cross sections $\sigma_{W}$
and $\sigma_{Z}$, and each corresponds to a particular direction
($L_{X}$) in the PDF parameter space (see Fig.\,\ref{fig:pedagogy}).
Therefore they are not suitable for estimating the PDF uncertainties
of other observables that are sensitive to other aspects of the PDFs.
Other PDF sets can be obtained, using the method introduced
in this paper, to probe the range of other variables,
such as rapidity ($y$) or transverse momentum ($p_{T} $)
distributions (hence relevant to the measurement of $W$ mass).
These will be investigated in subsequent work.
Also, the companion paper \cite{Hesse} supplies information
from the Hessian method that can be used to construct the
optimal PDFs for any observable $X$.

The PDF set that yields the ``best estimate'' for all of the physical
cross sections covered in this paper is our standard set $S_{0}$.
The parametrization of the initial distribution is given in
Appendix \ref{sec:AppPdfs}. In the following, we present
two sets of PDFs that bound the likely
range for each of the cross sections.

To exemplify the PDFs that characterize the range of uncertainty
of $W$ production at the Tevatron, we use two representative sets,
labeled $S_{W,\,{\rm TeV}}^{\pm }$, which correspond to
$\sigma_{W}=$ $\sigma_{W}(S_{0})\pm \Delta \sigma _{W}$
(with $\Delta \sigma _{W}/\sigma _{W} \sim 0.04$) respectively.
These two sets are extreme fits obtained by the Lagrange Multiplier
method.
The parameters $\{a\}$ for these sets are given
in Appendix \ref{sec:AppPdfs}.
We now compare some of the parton distributions from the three sets
($S_{W,\,{\rm TeV}}^{-},S_{0,\,}S_{W,\,{\rm TeV}}^{+}$),
to examine the ranges of variation of the PDFs themselves.

\figextremeW
Figure \ref{fig:extremeW}a shows $u(x,Q)$, $d(x,Q)$ and $g(x,Q)$
for $S_{W,\,{\rm TeV}}^{\pm }$, compared to the standard set $S_{0}$,
at $Q=80$\,GeV.
The function $x\Delta {f}(x)$ is plotted for each parton
flavor, where $\Delta {f}$ is $f-f_{0}$.
The gluon function has been divided by 10 to fit on the same graph.
The solid curves ($\Delta {u}_{+}$, $\Delta{d}_{+}$, $\Delta {g}_{+}$)
correspond to $S_{W,\,{\rm TeV}}^{+}$,
and the dashed curves to $S_{W,\,{\rm TeV}}^{-}$.
For $S_{W,\,{\rm TeV}}^{+}$, requiring $\sigma_{W}$ to be larger
than $\sigma^{(0)}_{W}$ makes the $u$ and $d$ distributions larger
than for the standard fit ($u_{0}$ and $d_{0}$) so $\Delta {u}_{+}$
and $\Delta{d}_{+}$ are positive.
Then the gluon distribution must be smaller
than the standard because of the momentum sum rule.
In the case of $S_{W,\,{\rm TeV}}^{-}$,
the reverse is true, resulting in almost a mirror behavior.
At the Tevatron, a typical $x$ for the parton-level process
$q_{1}\overline{q}_{2}\rightarrow W^{\pm }$ is $M_{W}/\sqrt{s}=0.04$.
The differences $\Delta{u}$ and $\Delta{d}$ are significant in the
range $0.01\leq x\leq 0.04$.
The magnitude of $\Delta{f}(x)$ in this range is a few percent of the
standard $f_{0}(x)$, which makes sense since $\Delta {\sigma }_{W}$
is a ${4}$\% shift of $\sigma_{W}$ for these PDF sets.

We can carry out the same comparison for $W$ production at the LHC.
The PDFs that bound the range of uncertainty are designated as
$S_{W,{\rm LHC}}^{\pm }$, which correspond to
$\sigma_{W}=\sigma_{W}(S_{0})\pm \Delta\sigma_{W}$
with $\Delta\sigma_{W}/\sigma_{W} \sim 0.08$.
The PDF parameters are given in Appendix \ref{sec:AppPdfs}.
Figure \ref{fig:extremeW}b shows parton distributions
from $S_{W,{\rm LHC}}^{\pm }$.
(Again, the gluon has been divided by 10.)
In the LHC case, the typical $x$ for the process
$q_{1}\bar{q}_{2}\rightarrow W^{\pm }$ is $M_{W}/\sqrt{s}=0.006$.
The region where $\Delta {u}$ and $\Delta {d}$ are significant is
seen accordingly lower in $x$ than for the Tevatron case.

In Appendix \ref{sec:AppPdfs} we also present PDF sets
$S_{Z,\,{\rm TeV}}^{\pm }$ and $S_{Z,{\rm LHC}}^{\pm }$
that characterize the range of uncertainties of $Z$ production
at the Tevatron and LHC.
These correspond to the outlying points on Figs.\ \ref{fig:Z0prod}a,b.
They are similar to
$S_{W,\,{\rm TeV}}^{\pm}$ and $S_{W,\,{\rm LHC}}^{\pm}$,
with small differences in the flavor dependence.

\section{Summary}
\label{sec:summary}

We have developed the Lagrange Multiplier method to calculate uncertainties
of physical observables due to PDFs, and we have used the method to
determine the uncertainty of the total cross sections for $W^{\pm }$
production and $Z^{0}$ production at the Tevatron and LHC.
The method is more reliable than past estimates because:
(i) it explores all the possibilities in the parameter space
of the input PDFs, independent of other assumptions; and (ii)
it produces the maximum allowed range for the specified physical
variables.
This is in contrast to previous attempts which relied on varying
certain features of the parton distributions chosen in some
{\it ad hoc} way.

From this analysis, we find that the uncertainty of
the prediction for $\sigma_{W}$ or $\sigma_{Z}$
at the Tevatron with current experimental constraints
is approximately $\pm{3}$--${4}$\%;
and at the LHC the uncertainties are approximately
$\pm{8}$--$10$\%.
These numbers do not include other uncertainties associated with
theoretical approximations, nuclear corrections, and other unexpected
sources. We have explored to some extent the possible effects due to the
choice of parametrization of the nonperturbative input PDFs, and found
them to be small.
The current work should be considered exploratory in nature,
as a first application of this improved approach to error
estimates.
A more comprehensive study, based on soon to be improved data sets,
and including other sources of uncertainties, will produce better
overall estimates of the physical predictions.

This study should be regarded as the precursor for many interesting
applications to come,
on physical processes of interest to the precision study of
the Standard Model, and on predictions for New Physics at future colliders.
Some examples are rapidity distributions of $W^{\pm }$ and $Z^{0}$
production, which contain a wealth of information on parton structure
of the nucleon; the $W$ mass measurement;
top and Higgs cross sections, {\it etc}.

There are other approaches to error estimates in global QCD
analysis \cite{ZomerDis,Alekhin,GieleEtal,Botje}.
In general, if greater emphasis is placed
on the ``rigor'' of the statistical method, then the range of
experiments that can be included in the analysis is narrower.
We have chosen to emphasize
the inclusion of the full range of experimental constraints, and adapt the
statistical analysis to deal with the practical problems that one
faces in such a system.
Within our general framework, there is an
alternative, complementary approach based on the conventional error matrix
method \cite{pdfuc0}.
We explore this latter method, as applied to global
QCD analysis of PDFs, in a companion paper \cite{Hesse}.
We mention briefly the contrasting features and relative
merits of the two approaches here.

The Lagrange Multiplier method focuses on a given physical observable $X$
(or a set of observables $\{X_{k}\}$) and determines the uncertainty $%
\Delta X$ allowed by the global data set within a specified tolerance for
the global fit.
The error matrix approach, using the Hessian matrix, focuses
instead on the uncertainties of the PDFs as represented by the parameters $%
\{a_{i}; i=1,\dots,d\}$.
It is in principle universal because, once determined,
these errors can be propagated to \emph{any} physical variable $X$.
However, the results are reliable
only if the function $\chi_{\rm global}^{2}(a)$
and the observable $X(a)$ can be approximated by lowest order
expansions in the parameters $\{a\}$,
and if the numerical computation of the derivatives
(the Hessian matrix) is under control.
The latter problem is surprisingly difficult
for global QCD analysis, because the eigenvalues
of the error matrix vary by many orders of magnitude.
This problem has been solved \cite{pdfuc0}, and
the error matrix results are consistent with the
constrained fitting results \cite{Hesse}.
Thus, at present, both methods appear to be applicable
to the study of uncertainties in global QCD analysis.

In Figures \ref{fig:WprodLHC} and \ref{fig:Z0prod}b
there is a significant cubic term in the dependence of
$\chi^{2}_{\rm global}$ on $\sigma_{W}$(LHC) and
$\sigma_{Z}$(LHC), respectively.
To calculate $\chi^{2}_{\rm global}$ versus $X$ accurately
in such cases, the Lagrange Multiplier method is necessary.
Traditional linear error analysis based on the Hessian
matrix can only produce a quadratic approximation to the
dependence.

When both methods are applicable, the Hessian method is more flexible and
easier to apply computationally.
But generally the Lagrange method is more robust and reliable.
As we expand the investigation to other physical processes of
interest, we will continue to test the efficacy of both methods
and cross check the results.

\paragraph*{Acknowledgement}
This work was supported in part by NSF grant PHY-9802564. We thank Michiel Botje
and Fabian Zomer for several interesting discussions and for valuable comments on
our work.  We also thank John Collins, Dave Soper, and other CTEQ colleagues
for discussions.

\appendix


\section{The effect of correlated errors
on $\mathbf{\Delta\chi^2}$}

\label{sec:AppDelChi}

The global fitting function $\chi^{2}_{\rm global}$
defined in (\ref{eq:Chi2global}) resembles the standard
statistical variable $\chi^{2}$, so it is tempting
to try to apply theorems of Gaussian statistics to
analyze the significance of the fit between theory
and experiment.
However, the familiar theorems do not apply, because
of correlations between measurement errors.
The purpose of this Appendix is to explore this issue.
The effect of correlated errors is potentially
a source of confusion.

For simplicity we describe the simplest case:
measurement of a single observable.
The arguments can be extended to cases where
multiple quantities are measured, such as the
determination of parton distribution functions.

Consider an observable $m$ that is measured $N$ times.
We shall refer to $N$ measurements of $m$ as one ``experiment''.
Let the true value of $m$ be $m_{0}$.
The measurements are $m_{1}, m_{2}, m_{3},\dots, m_{N}$.
The deviations from the true value are
$\alpha_{1}, \alpha_{2}, \alpha_{3},\dots, \alpha_{N}$,
where $\alpha_{i}=m_{i}-m_{0}$.
In general the measurement errors are correlated, so in the Gaussian
approximation the probability distribution of the
fluctuations is
\begin{equation}\label{eq:PD}
dP={\cal N}\exp\left\{-\frac{1}{2}\sum_{i,j=1}^{N}
\alpha_{i}C_{ij}\alpha_{j}\right\}d^{N}\alpha.
\end{equation}
Here $C_{ij}$ is a real symmetric matrix, and
${\cal N}=\sqrt{{\rm Det}\,C}/(2\pi)^{N/2}$ ensures
the normalization condition $\int dP=1$.

We will need the variance matrix
$\langle{\alpha_{i}\alpha_{j}}\rangle$, where
the notation $\langle{Q}\rangle$ means the average of
$Q$ in the probability distribution (\ref{eq:PD}).
For this Gaussian distribution,
\begin{equation}
\langle{\alpha_{i}\alpha_{j}}\rangle=\left(C^{-1}\right)_{ij}.
\end{equation}
The mean square fluctuation $E_{i}$ of the
$i^{\rm th}$ measurement $m_i$ is
\begin{equation}
E_{i} \equiv \langle\alpha_{i}^{2}\rangle
=\left(C^{-1}\right)_{ii}.
\end{equation}
To find the best estimate of the value of $m$ from these $N$ measurements,
{\em ignoring the correlations in the measurement errors}, we define
a chi-squared function $\chi_{u}^{2}(m)$ by
\begin{equation}\label{eq:defchi2}
\chi_{u}^{2}(m)=\sum_{i=1}^{N}\frac{\left(m_{i}-m\right)^{2}}{E_{i}}.
\end{equation}
The value of $m$ that minimizes $\chi_{u}^{2}(m)$, call it $\overline{m}$,
is then the best estimate of $m_{0}$ based on this information.
The function $\chi_{u}^{2}(m)$ is analogous to the fitting
function $\chi^{2}_{\rm global}$ in the CTEQ program, in the
sense that it does not include information about the
correlations between errors.
The minimum of $\chi_{u}^{2}(m)$ occurs at a weighted average of the
measurements,
\begin{equation}
\overline{m}=\frac{\sum_{i=1}^{N}m_{i}/E_{i}}{\sum_{i=1}^{N}1/E_{i}}.
\end{equation}
If all the $E_{i}$'s are equal then $\overline{m}$ is just the average
of the measurements.

Now, what are the fluctuations of the mean $\overline{m}$?
That is, if the ``experiment'' consisting of $N$ measurements
could be replicated many times, what would be the distribution
of $\overline{m}$'s obtained in those many trials?
It turns out that $\overline{m}$ has a Gaussian distribution
\begin{equation}
\frac{dP}{d\overline{m}}=\frac{1}{\sqrt{2\pi\Sigma^{2}}}
\exp\left[
-(\overline{m}-m_{0})^{2}/(2\Sigma^{2})\right].
\end{equation}
The standard deviation $\Sigma$ of $\overline{m}$ is
the RMS fluctuation; that is,
\begin{equation}
\Sigma^{2}
=\int\left(\overline{m}-m\right)^{2}\,dP
=\frac{1}{D^{2}}\sum_{ij}\frac{\left(C^{-1}\right)_{ij}}
{E_{i}E_{j}}
\end{equation}
where
\begin{equation}
D=\sum_{i}\frac{1}{E_{i}}.
\end{equation}

The question we wish to answer is this:
{\em How much does $\chi_{u}^{2}(m)$ increase, when $m$ moves
away from the minimum (at $\overline{m}$) by the amount $\pm \Sigma$
that corresponds to one standard deviation of the mean?}
The answer to this question is
\begin{equation}
\Delta{\chi_{u}^{2}}=\Sigma^{2}D.
\end{equation}
This result follows easily from the definition (\ref{eq:defchi2}),
because
\begin{equation}
\chi_{u}^{2}(\overline{m}+\Sigma) - \chi_{u}^{2}(\overline{m}) =
-2\Sigma\sum_{i}\frac{m_{i}-\overline{m}}{E_{i}}
+\Sigma^{2}\sum_{i}\frac{1}{E_{i}},
\end{equation}
and the term linear in $\Sigma$ is $0$ by the definition
of $\overline{m}$.
So far the discussion has been quite general.
We will now examine some illustrative special cases.

{\bf Example 1:} Suppose the measurement errors
are uncorrelated; that is,
\begin{equation}
C_{ij}=\delta_{ij}/E_{i}.
\end{equation}
Then the standard deviation of the mean $\overline{m}$ is
$\Sigma=1/\sqrt{D}$.
Thus for the uncorrelated case, the increase of $\chi_{u}^{2}$
corresponding to one standard deviation of the mean
is $\Delta{\chi_{u}^{2}}=1$.  This is the ``normal'' statistical
result: The $1\sigma$ range corresponds to an increase
of $\chi^{2}$ by 1.

An even more special case is when
the errors are uncorrelated and constant:
$E_{i}=\sigma^{2}$ independent of $i$,
where $\sigma$ is the standard deviation of
single measurements.  The correlation matrix
is $C_{ij}=\delta_{ij}/\sigma^{2}$.
In this case $D$ is $N/\sigma^{2}$, and the standard
deviation of the mean is $\Sigma=\sigma/\sqrt{N}$.

The criterion $\Delta{\chi}^{2}=1$ for one standard deviation
of a measured quantity is a standard result, often used
in the analysis of precision data.
But if $\chi^{2}$ is defined ignoring the correlations between
measurement errors, then the criterion $\Delta{\chi}^{2}=1$
is only valid for uncorrelated errors.
We will next consider two examples with correlated errors,
to show that $\Delta{\chi_{u}^{2}}$ is not $1$ for such cases.

{\bf Example 2:} Suppose measurements 1 and 2 are correlated,
3 and 4 are correlated, 5 and 6 are correlated, {\it etc.}
Then the correlation matrix is
\begin{equation}
C_{ij}=\left\{
\begin{array}{l}
1/\sigma^{2} {\rm ~for~} i=j
\\
c/\sigma^{2} {\rm ~for~} ij=12 {\rm ~or~} 21, 34 {\rm ~or~} 43, {\it etc}
\\
0 {\rm ~otherwise}\\
\end{array} \right.
\end{equation}
where $-1 < c < 1$ since the determinant of $C$ must be positive.
The inverse matrix $C^{-1}$ can be constructed using the
fact that $C$ is block diagonal, consisting of $N/2$
$2\times 2$ blocks.
Then it can be shown that
\begin{equation}
\Sigma=\frac{\sigma}{\sqrt{N}\sqrt{1+c}}
{\rm ~~and~~}
\Delta{\chi_{u}^{2}}=1-c.
\end{equation}
The increase of $\chi_{u}^{2}$ for one standard deviation
of the mean ranges from $0$ to $2$, depending on $c$.
The criterion $\Delta{\chi}^{2}=1$ does not apply
to this example with correlated errors.
A standard increase of $\chi_{u}^{2}$ may be smaller or larger than $1$.

{\bf Example 3:} For an even more striking example,
suppose the $N$ measurements that constitute a single
``experiment'' are, for $i=1, 2, 3,\dots, N$,
\begin{equation}
m_{i}=m_{0}+y_{i}+\beta
\end{equation}
where the $y_{i}$ are randomly distributed with standard
deviation $\sigma$,
and the measurements are systematically off by the amount $\beta$.
Suppose that $\beta$ has a Gaussian distribution
with standard deviation $s$ for replications of the
``experiment''.
In this example,
\begin{eqnarray}
C_{ij} &=& \frac{1}{\sigma^2}
\left(\delta_{ij} - \frac{s^2}{N s^2 + \sigma^2} \right),
\\
(C^{-1})_{ij} &=& \sigma^2 \, \delta_{ij} + s^2.
\end{eqnarray}
The variance of the individual measurements ($m_{i}$) is
\begin{equation}
\langle{m^{2}}\rangle-\langle{m}\rangle^{2}
=\sigma^{2} + s^{2}.
\end{equation}
Therefore our uncorrelated chi-squared variable
$\chi_{u}^{2}(m)$, defined ignoring the correlations, is
\begin{equation}
\chi_{u}^{2}(m)=\sum_{i=1}^{N}\frac{\left(m-m_{i}\right)^{2}}
{\sigma^{2} + s^{2}}.
\end{equation}
The minimum of $\chi_{u}^{2}(m)$ occurs at $\overline{m}$, which is just the
average of the individual measurements.
The variance of $\overline{m}$, averaged over many replications
of the ``experiment'', is
\begin{equation}
\Sigma^{2}=\langle{\overline{m}^{2}}\rangle -\langle{\overline{m}}\rangle^{2}
          =s^{2} + \frac{\sigma^{2}}{N}.
\end{equation}
The increase of $\chi_{u}^{2}$ as $m$ moves from $\overline{m}$
to $\overline{m}\pm\Sigma$, {\it i.e.}, by one standard deviation
of the mean, is
\begin{equation}
\Delta{\chi_{u}^{2}} \equiv
\chi_{u}^{2}(\overline{m}+\Sigma)-\chi_{u}^{2}(\overline{m})
 = \frac{\sigma^{2} + N s^{2}}{\sigma^{2} + s^{2}}.
\end{equation}
In the limit $s/\sigma \ll 1$, the error correlations in this
model become negligible and
$\Delta\chi^{2}$ reduces to the conventional value of $1$.
But in the limit  $s/\sigma \gg 1$ where the error correlations
are dominant, $\Delta\chi^{2}$ approaches $N$.

Thus for Example 3---a systematic error with
100\% correlation between measurements---the
increase of $\chi_{u}^{2}$ for a standard deviation
of $\overline{m}$ is much larger than 1.
If $s$ and $\sigma$ are comparable, then $\Delta{\chi_{u}^{2}}$
is of order $N$.

If the correlation matrix $C_{ij}$ is known accurately,
then the correlation information can be incorporated
into the definition of the $\chi^{2}$ function,
in the manner of Appendix \ref{sec:AppCorSys}.
For the full list of experiments in the global analysis
of parton distribution functions, however, the correlations of
systematic errors have not been published, so the fitting
function $\chi^{2}_{\rm global}$ has only uncorrelated
systematic errors.

We described above the measurement of a single quantity.
The determination of parton distribution functions seeks
to measure {\em many} quantities, {\it i.e.,} the
$16$ parameters $\{a\}$.
The above arguments can be extended to measurements of
multiple quantities.
If the measurement errors are uncorrelated, then the
increase of $\chi^{2}_{u}$ by 1 from the minimum defines
a hyperellipse in parameter
space---the {\em error ellipse}---corresponding
to one standard deviation of linear combinations
of the parameters.
However, if the errors are correlated then $\Delta{\chi_{u}^{2}}=1$
is not the correct criterion for a standard deviation.

The Lagrange Multiplier method finds the best fit to
the data, subject to a constrained value of some quantity $X$.
The prediction of $X$ is at the absolute minimum of $\chi^{2}_{u}$.
Again, if the errors are uncorrelated then one standard
deviation of the constrained quantity corresponds to
an increase of $\chi^{2}$ by 1 from the absolute minimum.
But if the errors are correlated then $\Delta{\chi_{u}^{2}}=1$
is not the correct criterion for one standard deviation
of $X$.

One reason for describing this familiar, even elementary,
statistics, is to avoid certain misconceptions.
Our standard PDF set $S_{0}$
is a parametrized fit to 1295 data points
with 16 fitting parameters.
The minimum value of $\chi^{2}_{\rm global}$ is approximately
1200.
Naively, it seems that an increase of $\chi^{2}_{\rm global}$
by merely 1, say from 1200 to 1201, could not possibly
represent a standard deviation of the fit.
Naively one might suppose that a standard deviation
would have $\Delta{\chi}^{2}\sim\sqrt{1295}$ rather than 1.
However, this is a misconception.
If the errors are uncorrelated
(or if the correlations are incorporated into
$\chi^{2}$) then indeed $\Delta{\chi}^{2}=1$ would
represent a standard deviation.
But this theorem is irrelevant to our problem,
because the large correlations of systematic errors
are not taken into account in $\chi^{2}_{\rm global}$.

\section{\protect$\mathbf{\chi^2}$
function including correlated systematic errors}

\label{sec:AppCorSys}

The purpose of this appendix is to derive the appropriate definition
of $\chi^{2}$ for data with correlated systematic errors.
The defining condition is that $\chi^{2}$ should obey a
chi-squared distribution.

Let $\left\{m_{i}\right\}$ be a set of measurements, where
$i=1, 2, 3, \dots, N$.
Let $t_{i}$ be the true, {\it i.e.,} theoretical value
of the $i^{\rm th}$ measured quantity.
Several kinds of measurement errors will
contribute to the difference between $m_{i}$ and $t_{i}$.
The uncorrelated error of measurement $i$
is denoted by $\sigma_{i}$.
There are also correlated errors, $K$ in number,
denoted $\beta_{1i}, \beta_{2i}, \dots, \beta_{Ki}$.
Thus the $i^{\rm th}$ measurement can be written as
\begin{equation} \label{eq:mistper}
m_{i} = t_{i}+{\rm ~errors}
 = t_{i}+\sigma_{i}r_{i}+\sum_{j=1}^{K}\beta_{ji}r_{j}^{\prime}
\end{equation}
where $r_{i}$ and $r_{j}^{\prime}$ are independently
fluctuating variables.
We assume that each of these fluctuations has a Gaussian distribution
with width $1$,
\begin{equation}
p(r)=\frac{e^{-r^{2}/2}}{\sqrt{2\pi}}.
\end{equation}
Note that $r_{j}^{\prime}$ is independent of $i$;
that is, the errors $\beta_{j1}, \beta_{j2},\dots ,\beta_{jN}$
are 100\% correlated for all $N$ data points.

The probability distribution of the measurements is
\begin{eqnarray}
dP &=& \int \prod_{i=1}^{N} p(r_{i})dr_{i}
\prod_{j=1}^{K} p(r_{j}^{\prime}) dr_{j}^{\prime}
\nonumber \\
 &\times& \prod_{i=1}^{N}\delta
\left(m_{i}-t_{i}-\sigma_{i}r_{i}
-\sum_{j=1}^{K}\beta_{ji}r_{j}^{\prime}\right)d^{N}m.
\label{eq:PDwithC}
\end{eqnarray}
Now we will evaluate the integrals over $r_{i}$ and $r^{\prime}_{j}$
in two steps.
First evaluate the $r_{i}$ integrals using the delta functions,
\begin{equation}
dP = \int \prod_{j=1}^{K} dr^{\prime}_{j} {\cal C}_{1}
e^{-\chi_{1}^{2}/2} \;d^{N}m
\end{equation}
where ${\cal C}_{1}$ is a normalization constant and
\begin{equation}
\chi_{1}^{2}=
\sum_{i=1}^{N} \left(\frac{m_{i}-t_{i}
-\sum_{j}\beta_{ji}r^{\prime}_{j}}{\sigma_{i}}\right)^{2}
+\sum_{j=1}^{K} {r^{\prime}_{j}}^{2}.
\end{equation}
Note that $\chi_{1}^{2}$ is a function of $r_{1}^{\prime},
\dots,r_{K}^{\prime}$.
These variables $\{r_{j}^{\prime}\}$ could be used as fitting
parameters to account for the systematic errors:
Minimizing $\chi_{1}^{2}$ with respect to $r_{j}^{\prime}$ would
provide the best model to correct for the systematic error of
type $j$.
Because $\chi_{1}^{2}$ is only a quadratic polynomial in
the $r_{j}^{\prime}$ variables, the minimization can be
done analytically.

To continue evaluating (\ref{eq:PDwithC}) we now do the integration
over $\{r_{j}^{\prime}\}$.
Write $\chi_{1}^{2}$ in the form
\begin{equation}
\chi_{1}^{2} = \sum_{i=1}^{N}\frac{\left(m_{i}-t_{i}\right)^{2}}
{\sigma_{i}^{2}} - \sum_{j=1}^{K} 2 B_{j} r_{j}^{\prime}
+\sum_{j,j^{\prime}=1}^{K}A_{jj^{\prime}}r_{j}^{\prime}
r_{j^{\prime}}^{\prime}
\end{equation}
where $B_{j}$ is a vector with $K$ components
\begin{equation}
B_{j}=\sum_{i=1}^{N}\beta_{ji}(m_{i}-t_{i})/\sigma_{i}^{2},
\end{equation}
and $A_{jj^{\prime}}$ is a $K\times K$ matrix
\begin{equation}
A_{jj^{\prime}}=\delta_{jj^{\prime}}
+\sum_{i=1}^{N}\beta_{ji}\beta_{j^{\prime}i}/\sigma_{i}^{2}.
\end{equation}
Then the integration over $d^{K}r^{\prime}$ is an
exercise in Gaussian integration, with the result
\begin{equation}
dP={\cal C}\exp\left[-\frac{1}{2}\chi^{2}\right] d^{N}m
\end{equation}
where ${\cal C}$ is a normalization constant and
\begin{equation}\label{eq:th-data_corr}
\chi^{2}=\sum_{i=1}^{N}\frac{(m_{i}-t_{i})^{2}}{\sigma_{i}^{2}}
-\sum_{j=1}^{K}\sum_{j^{\prime}=1}^{K}
B_{j}\left(A^{-1}\right)_{jj^{\prime}}B_{j^{\prime}}.
\end{equation}
This equation is the appropriate definition of $\chi^{2}$
for data with correlated systematic errors.
The correlated errors are defined by the coefficients
$\beta_{ji}$ in (\ref{eq:mistper}), which determine the
vector $B_{j}$ and matrix $A_{jj^{\prime}}$.
An interesting relation is that the $\chi^{2}$ quantity in
(\ref{eq:th-data_corr}) is the minimum of $\chi_{1}^{2}$
with respect to the parameters
$r_{1}^{\prime},\dots,r_{K}^{\prime}$.

Another expression for $\chi^{2}$, which may be derived
from (\ref{eq:PDwithC}) by Gaussian integration, is \cite{Alekhin}
\begin{equation}\label{eq:Alekhin}
\chi^{2}=\sum_{i=1}^{N}\sum_{i^{\prime}=1}^{N}
\left(m_{i}-t_{i}\right) \left(V^{-1}\right)_{ii^{\prime}}
\left(m_{i^{\prime}}-t_{i^{\prime}}\right)
\end{equation}
where $V_{ij}$ is the variance matrix
\begin{equation}
V_{ii^{\prime}}=\sigma_{i}^{2} \delta_{ii^{\prime}}
+\sum_{j=1}^{K} \beta_{ji} \beta_{j i^{\prime}}.
\end{equation}
It can be shown that the inverse of the variance matrix is
\begin{equation}
\left(V^{-1}\right)_{ii^{\prime}}=
\frac{\delta_{ii^{\prime}}}{\sigma_{i}^{2}}-\sum_{j,j^{\prime}=1}^{K}
\frac{\beta_{ji}\beta_{j^{\prime}i^{\prime}}}
{{\sigma_{i}}^{2}{\sigma_{i^{\prime}}^{2}}}
\left(A^{-1}\right)_{jj^{\prime}}.
\end{equation}
Therefore (\ref{eq:th-data_corr})
and (\ref{eq:Alekhin}) are equivalent.
However, there is a real computational advantage in the use of
(\ref{eq:Alekhin}) because it does not require the numerical
inversion of the $N\times{N}$ variance matrix.

To check that (\ref{eq:th-data_corr}) makes sense
we can consider a special case.
Suppose the number $K$ of systematic errors is $N$,
and each systematic error contributes to just one
measurement.
Then the matrix of systematic errors has the form
\begin{equation}
\beta_{ji}=\delta_{ji}b_{i}.
\end{equation}
This situation is equivalent to an additional set of
{\em uncorrelated} errors $\left\{b_{i}\right\}$.
The vector $B_{j}$ is then
\begin{equation}
B_{j}=\frac{b_{j}\left(m_{j}-t_{j}\right)}{\sigma_{j}^{2}}
\end{equation}
and the matrix $A_{jj^{\prime}}$ is
\begin{equation}
A_{jj^{\prime}}=\delta_{jj^{\prime}}\left[1+\frac{b_{j}^{2}}
{\sigma_{j}^{2}}\right].
\end{equation}
Substituting these results into (\ref{eq:th-data_corr}) we find
\begin{equation}
\chi^{2}=\sum_{i}\frac{\left(m_{i}-t_{i}\right)^{2}}{\sigma_{i}^{2}
+b_{i}^{2}},
\end{equation}
which makes sense:
the uncorrelated errors just combine in quadrature.

The statistical quantity $\chi^{2}$ has a chi-squared
distribution with $N$ degrees of freedom.
Thus this variable may be used to set confidence levels
of the theory for the given data.
But to use this variable, the measurement errors
$\sigma_{i}$ and $\beta_{ji}$, for
$i=1,2,\dots,N$ and $j=1,2,\dots,K$, must be known
from the experiment.
A chi-squared distribution with many degrees of freedom
is a very narrow distribution, sharply peaked at
$\chi^{2}\!=\!N$.
Therefore small inaccuracies in the values of the
$\sigma_{i}$'s and $\beta_{ji}$'s may translate into
a large error on the confidence levels computed from
the chi-squared distribution.

It is equation (\ref{eq:th-data_corr}) that we use
in Section \ref{sec:quantifying} to compare the constrained
fits produced by the Lagrange multiplier method to data from
the H1 and BCDMS experiments.
Correlated systematic errors are also used to calculate
$\chi^{2}$ for the CDF and D0 jet experiments.

%

\section{Parton Distribution Sets}

\label{sec:AppPdfs}

We give here the PDFs described in Section 6.
$S_{0}$ is the standard set, defined by the absolute minimum
of $\chi^{2}_{\rm global}$.
$S^{\pm}_{W,{\rm Tev}}$ are fits to the global data sets
with extreme values of $\sigma_{W}$(Tevatron),
{\it i.e.,} the outermost points on Fig.\ \ref{fig:Wprod},
generated by the Lagrange multiplier method.
$S^{\pm}_{Z,{\rm Tev}}$, $S^{\pm}_{W,{\rm LHC}}$, and
$S^{\pm}_{Z,{\rm LHC}}$ are analogous for $Z$ production
and $W$ and $Z$ production at the LHC.

\bigskip

The functional form of the initial parton distributions and the
definitions of the PDF parameters at the low-energy scale
$Q_{0}=1$\,GeV are
\[
f(x,Q_{0}^{2})=A_{0}\,x^{A_{1}}(1-x)^{A_{2}}(1+A_{3}\,x^{A_{4}})
\]
for $u_{v},d_{v},g,\bar{u}+\bar{d},s(=\bar{s})$; and for the ratio
\[
\frac{\bar{d}(x,Q_{0}^{2})}
{\bar{u}(x,Q_{0}^{2})}
=A_{0}\,\,x^{A_{1}}(1-x)^{A_{2}}+(1+A_{3}\,x)\,(1-x)^{A_{4}}.
\]
The tables of coefficients follow.

\[
S_{0}: \quad
\begin{array}{|c|r|r|r|r|r|}
\hline
      & A_{0}   & A_{1}   & A_{2}   & A_{3}   & A_{4}  \\  \hline
d_{v} & 0.5959  & 0.4942  & 4.2785  & 8.4187  & 0.7867 \\  \hline
u_{v} & 0.9783  & 0.4942  & 3.3705  & 10.0012 & 0.8571 \\  \hline
    g & 3.3862  & 0.2610  & 3.4795  & -0.9653 & 1. \\  \hline
\bar{d}/\bar{u}
      & 3.051E4 & 5.4143  & 15. & 9.8535  & 4.3558 \\  \hline
\bar{u}+\bar{d} 
      & 0.5089  & 0.0877  & 7.7482  & 3.3890  & 1. \\  \hline
    s & 0.1018  & 0.0877  & 7.7482  & 3.3890  & 1. \\  \hline
\end{array}
\]

\bigskip

\[
S_{W,\mathrm{TeV}}^\pm: \quad
\begin{array}{|c|r|r|r|r|r|}\hline
 & A_{0} & A_{1} & A_{2} & A_{3} & A_{4} \\ \hline
d_v
 & 0.2891 & 0.5141 & 3.8555 & 10.9580 & 0.4128 \\ 
 & 0.2184 & 0.2958 & 4.6267 & 35.7229 & 1.0958 \\ \hline
u_v
 & 1.0142 & 0.5141 & 3.3614 & 9.2995 & 0.8053 \\ 
 & 0.2979 & 0.2958 & 3.3279 & 32.8453 & 0.9427 \\ \hline
g
 & 4.6245 & 0.4354 & 3.4795 & -0.9728 & 1. \\ 
 & 1.8080 & 0.0458 & 3.4795 & -0.0519 & 1. \\ \hline
\bar{d}/\bar{u}
 & 5.908E4 & 5.6673 & 15. & 9.8535 & 4.7458 \\ 
 & 2.041E4 & 5.1506 & 15. & 9.8535 & 4.8320 \\ \hline
\bar{u}+\bar{d}
 & 0.4615 & 0.0108 & 6.6145 & 0.92784 & 1. \\ 
 & 1.2515 & 0.3338 & 7.5216 & -0.0570 & 1. \\ \hline
s
 & 0.0923 & 0.0108 & 6.6145 & 0.9278 & 1. \\ 
 & 0.2503 & 0.3338 & 7.5216 & -0.0570 & 1. \\ \hline
\end{array}
\]

\bigskip

\[
S_{Z,\mathrm{TeV}}^\pm: \quad
\begin{array}{|c|r|r|r|r|r|} \hline
 & A_{0} & A_{1} & A_{2} & A_{3} & A_{4} \\ \hline
d_v
 & 0.6061 & 0.5502 & 4.0017 & 5.8346 & 0.5343 \\ 
 & 0.3427 & 0.3728 & 4.5166 & 19.8510 & 0.9966 \\ \hline
u_v
 & 1.2159 & 0.5502 & 3.3347 & 7.3386 & 0.7711 \\ 
 & 0.5247 & 0.3728 & 3.3905 & 20.1006 & 0.9556 \\ \hline
g
 & 4.4962 & 0.4321 & 3.4795 & -0.9023 & 1. \\ 
 & 2.3113 & 0.1032 & 3.4795 & -0.6349 & 1. \\ \hline
\bar{d}/\bar{u}
 & 4.321E4 & 5.4724 & 15. & 9.8535 & 4.6298 \\ 
 & 2.818E4 & 5.4540 & 15. & 9.8535 & 4.4376 \\ \hline
\bar{u}+\bar{d}
 & 0.4609 & 0.0103 & 6.6671 & 0.9822 & 1. \\ 
 & 0.9900 & 0.2926 & 8.3205 & 2.1648 & 1. \\ \hline
s
 & 0.0921 & 0.0103 & 6.6671 & 0.9823 & 1. \\ 
 & 0.1980 & 0.2926 & 8.3205 & 2.1648 & 1. \\ \hline

\end{array}
\]

\bigskip

\[
S_{W,\mathrm{LHC}}^\pm: \quad
\begin{array}{|c|r|r|r|r|r|} \hline
 & A_{0} & A_{1} & A_{2} & A_{3} & A_{4} \\ \hline
d_v
 & 0.7326 & 0.5008 & 4.6393 & 10.8532 & 1.0595 \\ 
 & 0.5671 & 0.4771 & 4.2615 &  8.8355 & 0.8130 \\ \hline
u_v
 & 1.0608 & 0.5008 & 3.4023 &  9.6622 & 0.8968 \\ 
 & 0.9142 & 0.4771 & 3.3761 & 10.9138 & 0.8809 \\ \hline
g
 & 2.2379 & 0.0733 & 3.4795 & -0.9860 & 1. \\ 
 & 2.5021 & 0.3981 & 3.4795 &  1.6229 & 1. \\ \hline
\bar{d}/\bar{u}
 & 2.178E4 & 5.2576 & 15. & 9.8535 & 4.4810 \\ 
 & 4.531E4 & 5.4979 & 15. & 9.8535 & 4.6585 \\ \hline
\bar{u}+\bar{d}
 & 1.1980 &  0.2952 & 6.9475 & -0.5442 & 1. \\ 
 & 0.2759 & -0.0918 & 8.2045 &  6.3950 & 1. \\ \hline
s
 & 0.2396 &  0.2952 & 6.9475 & -0.5442 & 1. \\ 
 & 0.0552 & -0.0918 & 8.2045 &  6.3950 & 1. \\ \hline

\end{array}
\]

\bigskip

\[
S_{Z,\mathrm{LHC}}^\pm: \quad
\begin{array}{|c|r|r|r|r|r|} \hline

 & A_{0} & A_{1} & A_{2} & A_{3} & A_{4} \\ \hline
d_v
 & 0.5659 & 0.4616 & 4.5297 & 12.3685 & 0.9836 \\ 
 & 0.4585 & 0.4496 & 4.2122 & 10.3850 & 0.7760 \\ \hline
u_v
 & 0.8344 & 0.4616 & 3.3847 & 12.1129 & 0.8872 \\ 
 & 0.7640 & 0.4496 & 3.3566 & 12.8253 & 0.8701 \\ \hline
g
 & 2.3282 & 0.0918 & 3.4795 & -0.9837 & 1. \\ 
 & 2.9475 & 0.4219 & 3.4795 & 0.9447 & 1. \\ \hline
\bar{d}/\bar{u}
 & 2.421E4 & 5.3032 & 15. & 9.8535 & 4.5341 \\ 
 & 4.416E4 & 5.4708 & 15. & 9.8535 & 4.7925 \\ \hline
\bar{u}+\bar{d}
 & 1.1130 &  0.2698 & 6.8490 & -0.5330 & 1. \\ 
 & 0.2719 & -0.0899 & 8.1492 &  6.5300 & 1. \\ \hline
s
 & 0.2226 & 0.2698 & 6.8490 & -0.5330 & 1. \\ 
 & 0.0544 & -0.0899 & 8.1492 & 6.5300 & 1. \\ \hline

\end{array}
\]
With a program to solve the PDF evolution equations,
the PDFs for an arbitrary momentum scale $Q$ can be
generated.

\newpage


\end{document}